\documentclass[conference]{IEEEtran}
\usepackage[T1]{fontenc}
\usepackage{booktabs}
\usepackage{balance}
\usepackage{import}
\usepackage{caption}
\usepackage{subcaption}
\usepackage{xcolor}
\usepackage{xspace}
\usepackage{graphicx}
\graphicspath{{picture/}}
\usepackage{multirow}
\usepackage{tabularx}
\usepackage{makecell}
\usepackage{hyperref}
\usepackage{comment}
\usepackage{caption}
\usepackage{amsmath}
\usepackage{enumitem}
\usepackage{amsmath}
\usepackage{amsfonts}  % or \usepackage{amssymb}
\usepackage{cite}

\newcommand{\pname}{\textbf{\texttt{ARS}}\xspace}

\begin{document}

% \title{Ambient RF Sensing: Leveraging Over-the-Air  5G Radio Signals for Human Activity Detection}
\title{Spectrum Shortage for Radio Sensing? Leveraging Ambient 5G Signals for Human Activity Detection}

% \renewcommand{\shortauthors}{XXx et al.}
% \renewcommand{\shorttitle}{Ambient RF Sensing}

%%
%% The code below is generated by the tool at http://dl.acm.org/ccs.cfm.
% \begin{CCSXML}
% <ccs2012>
%    <concept>
%        <concept_id>10010520.10010553.10010559</concept_id>
%        <concept_desc>Computer systems organization~Sensors and actuators</concept_desc>
%        <concept_significance>500</concept_significance>
%        </concept>
%    <concept>
%        <concept_id>10010147.10010178.10010224</concept_id>
%        <concept_desc>Computing methodologies~Computer vision</concept_desc>
%        <concept_significance>500</concept_significance>
%        </concept>
%  </ccs2012>
% \end{CCSXML}

% \ccsdesc[500]{Computer systems organization~Sensors and actuators}
% \ccsdesc[500]{Computing methodologies~Computer vision}

\author{
\IEEEauthorblockN{Kunzhe Song, Maxime Zingraff, and Huacheng Zeng}
\IEEEauthorblockA{Department of Computer Science and Engineering, Michigan State University, USA}
}

\maketitle
\begin{abstract}
Radio sensing in the sub-10 GHz spectrum offers unique advantages over traditional vision-based systems, including the ability to see through occlusions and preserve user privacy. However, the limited availability of spectrum in this range presents significant challenges for deploying large-scale radio sensing applications. In this paper, we introduce Ambient Radio Sensing (\pname), a novel Integrated Sensing and Communications (ISAC) approach that addresses spectrum scarcity by repurposing over-the-air radio signals from existing wireless systems (e.g., 5G and Wi-Fi) for sensing applications, without interfering with their primary communication functions. \pname operates as a standalone device that passively receives communication signals, amplifies them to illuminate surrounding objects, and captures the reflected signals using a self-mixing RF architecture to extract baseband features. This hardware innovation enables robust Doppler and angular feature extraction from ambient OFDM signals. 
To support downstream applications, we propose a cross-modal learning framework focusing on human activity recognition, featuring a streamlined training process that leverages an off-the-shelf vision model to supervise radio model training. We have developed a prototype of \pname and validated its effectiveness through extensive experiments using ambient 5G signals, demonstrating accurate human skeleton estimation and body mask segmentation applications.

% This work introduces a scalable, privacy-preserving, and spectrum-efficient radio sensing paradigm, offering a promising pathway for future ISAC systems in smart healthcare, automation, and surveillance applications.

\end{abstract}

% \begin{IEEEkeywords}
%     5G, integrated sensing and communications (ISAC), human activity detection, radio spectrum
% \end{IEEEkeywords}


\section{Introduction}
 
Radio sensing on sub-10 GHz spectrum bands offers several compelling advantages over camera and LiDAR sensors, particularly in challenging environments. Unlike cameras, radio waves are not affected by lighting conditions, allowing reliable operation in darkness, fog, or glare. Compared to LiDAR, radio sensing is more robust in rain, dust, and smoke, and can penetrate obstacles like walls or foliage, enabling non-line-of-sight detection. Additionally, radio sensing systems are privacy-preserving for human monitoring applications such as nursing homes, elderly care, or hospitals, as they can detect presence, movement, and vital signs without capturing identifiable visual information like faces or body details. This makes them ideal for continuous, non-intrusive monitoring while respecting individuals' privacy.

While radio sensing technologies such as FMCW radar have become very mature and cost-effective, their real-world applications remain limited. 
The primary obstacle lies in the scarcity of spectrum resources. 
The sub-10 GHz spectrum has been intensively allocated or reserved for many important applications such as TV broadcasting, cellular networks (e.g., 4G/5G), satellite communications, Wi-Fi, military and aviation radar, and public safety systems. As a result, introducing new sensing services in this spectrum range faces strict regulatory constraints and interference challenges, limiting the deployment of radio sensing technologies at scale.

One approach to addressing the spectrum scarcity is to use high-frequency bands, such as the millimeter-wave (mmWave) spectrum, for sensing purposes.
This approach has been widely adopted, leading to the development of a variety of mmWave FMCW radars for emerging applications such as autonomous driving, and health monitoring. %, and industrial automation.
The abundant bandwidth and short wavelengths of mmWave frequencies enable high-resolution sensing, making them well-suited for fine-grained environmental perception.
Despite their success in various applications, mmWave radars have very limited obstacle-penetration capabilities and a short detection range.
%, high power consumption, and high hardware complexity. 

\begin{figure}
    \centering
    \includegraphics[width=1\linewidth]{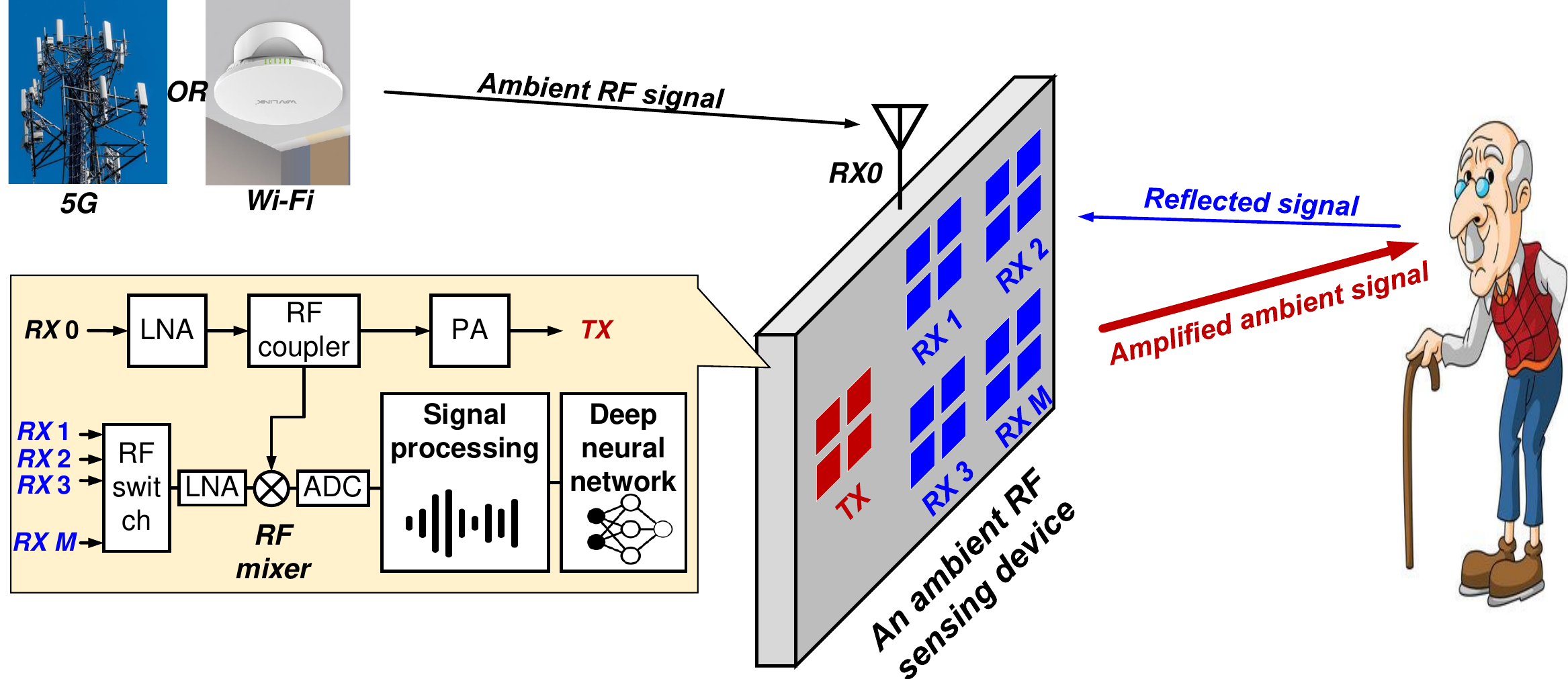}
    \caption{Architectural diagram of \pname. 
    % \pname receives over-the-air radio signals from an existing wireless communication system (e.g., 5G or Wi-Fi), amplifies them, and re-transmits the signals to illuminate surrounding objects. Simultaneously, it receives and demodulates the signals reflected from moving objects to extract heatmap features. A deep neural network (DNN) is then employed to perform downstream tasks such as human skeleton and body mask estimations.
    }
    \label{fig:ambientRFsensing}
\end{figure}

\begin{comment}
Another approach to addressing the spectrum scarcity issue is to integrate radio sensing into existing wireless communication systems by leveraging received signal strength (RSS), channel state information (CSI), or other readily available signal features.
In particular, CSI-based sensing has been extensively studied in Wi-Fi networks, enabling increasingly sophisticated techniques for human activity recognition \cite{yousefi2017survey, zou2018deepsense, li2021two, zou2019wifi, xue2020deepmv, xiao2020deepseg, sheng2020deep, schafer2021human, moshiri2021csi, 9726794, yang2022efficientfi, yang2022autofi, zhang2018crosssense, gu2021wione, yang2022securesense, zou2018robust, yang2019learning, xiao2019csigan, li2020wihf, wang2022airfi, gu2022wigrunt, zhang2021wifi, zhang2021widar3}.
While this approach does not increase spectrum utilization for sensing, it has some fundamental limitations in its sensing performance. 
First, the CSI-based sensing approach is not capable of extracting Doppler signatures of moving objects due to the lack of synchronization between communication transmitter and receiver devices \cite{song2024siwis}. 
As Doppler signatures are key to inferring the moving pattern of objects, the CSI-based sensing approach appears degraded performance when generalized into new scenarios. 
Second, and more importantly, CSI is only available within the communication systems, making it impossible to deploy a sensing device at the locations where access to the communication systems is restricted or unavailable.
\end{comment}

In this paper, we propose a new approach for sub-10GHz radio sensing to address the spectrum scarcity issue. 
Our approach is based on the fact that radio waves from wireless communication systems such as 5G and Wi-Fi are ubiquitous, both indoors and outdoors. 
The key idea is to leverage the over-the-air radio waves from existing communication systems for radio sensing applications while not negatively affecting communication system performance. 
To validate this approach, we introduce an Ambient Radio Sensing (\pname) device, as shown in Fig.~\ref{fig:ambientRFsensing}.
\pname is a standalone radio sensing device. 
On the transmitter side, it receives the over-the-air radio signals from an existing communication system, such as a 5G base station, and amplifies them to illuminate surrounding objects for sensing purpose.
On the receiver side, it captures the reflected signals from the environment and down-converts them to the baseband for feature extraction and learning-based object detection.
Crucially, since \pname adopts an amplify-and-forward approach for RF signal processing in the analog domain, it does not generate interference to existing communication systems. Instead, it potentially boosts the signal strength for the communication system.
% We stress that, since \pname adopts the amplify-and-forward approach for the RF signal process in the analog domain, it does not generate any interference to the existing communication system. 
% Instead, it boosts the signal strength for the communication system. 

% \pname comprises four main components: 
% (i) dipole and patch antennas, (ii) RF circuits, (iii) signal processing blocks for feature extraction, and (iv) deep learning modules for downstream tasks. 
% Combining the hardware and algorithmic innovations, it is capable of extracting reliable temporal-coherence features for downstream sensing applications. 
On the hardware side, we propose a self-mixing RF architecture to generate baseband signals optimized for feature extraction. 
Specifically, the dipole antenna (labeled “RX0” in Fig.~\ref{fig:ambientRFsensing}) receives over-the-air radio signals from a specific communication system and amplifies them to illuminate nearby objects using a patch antenna (labeled “TX” in Fig.~\ref{fig:ambientRFsensing}), which is highly optimized for the specific communication spectrum band and serving as a band-pass filter to suppress the signals on unintended spectrum bands. 
% This amplify-and-forward operation does not generate interference to the communication devices but instead serves as a relay to boost signal strength for them.
In parallel, multiple patch antennas (labeled “RX1” to “RX\;$M$” in Fig.~\ref{fig:ambientRFsensing}) are employed to receive the reflected signals from surrounding objects for sensing purposes.
These reflected signals are first amplified and then mixed with a copy of the amplified signal from the dipole antenna, generating a baseband signal for digitization and feature extraction.
%These reflected signals are first amplified and then mixed with a copy of the amplified signal from the dipole antenna, generating a baseband signal for digitization and feature extraction.
This innovative self-mixing RF architecture enables \pname to extract coherent temporal and spatial features crucial for motion analysis.
%thereby approaching the capability and performance of a monostatic radar.

On the algorithmic side, the OFDM waveform and limited bandwidth of existing signals present significant challenges for feature extraction.
A key question is whether it is possible to extract Doppler signatures of moving objects from the baseband signal generated by our sensing device.
Our answer is affirmative.
Through analytical studies of 5G NR OFDM frames, we find that, through proper signal processing, the phase of the generated baseband signal exhibits an approximately linear relationship with object displacement.
This deterministic relationship enables continuous estimation of Doppler signatures. % of moving objects, which have been shown to be effective features for modeling motion patterns in both model-based and learning-based downstream tasks.
Additionally, by leveraging multiple receiving antennas, \pname can estimate the angular direction of these signatures, providing a rich spatio-temporal representation for downstream tasks.

To demonstrate the potential of \pname, we implement two demanding downstream tasks: human skeleton estimation and body mask segmentation. Since radio features are inherently sparse and sensitive to environmental noise, capturing fine-grained human poses is challenging. We address this by adopting a cross-modal supervised learning approach, which distills high-fidelity knowledge from the vision domain into the radio domain. 
During training, an off-the-shelf vision model provides ground-truth labels from synchronized video to supervise the radio-based DNN. We have validated this framework through a functional prototype and extensive experiments, achieving state-of-the-art performance on both tasks.
% Specifically, during training, we equip \pname with a video camera to provide ground-truth labels, which are used to supervise the learning of the DNN for radio-based activity recognition.
% Once trained, the model performs well on both tasks using only radio signal features as input.

% We have built a prototype of \pname and evaluated its performance for human skeleton and body mask estimations. 
% Extensive experiments show that \pname achieves state-of-the-art performance on both tasks. 

The primary contributions of this work are as follows:
\begin{itemize}
\item 
It presents a novel approach to Integrated Sensing and Communications (ISAC) that addresses spectrum scarcity by repurposing ubiquitous ambient signals.

\item 
% It introduces a comprehensive solution that combines joint hardware and algorithmic designs to enable reliable feature extraction for human activity detection.
It introduces a joint hardware-algorithmic design featuring a self-mixing RF architecture for robust feature extraction from ambient OFDM waveforms.

\item 
% It demonstrates the effectiveness of our approach through human skeleton and body mask estimation tasks. Extensive experiments confirm its practicality and superior performance in real-world scenarios.
It validates the practicality of ARS through extensive experiments, demonstrating superior performance in real-world human skeleton and mask estimation scenarios.
\end{itemize}
% \import{section}{preliminary}

\section{Radar-Inspired RF Hardware}

% \subsection{Overview}
\begin{comment}
    
Given the spectrum scarcity faced by radar devices and the limitation of CSI-based sensing approaches, we aim to design a new paradigm for radio using ambient RF signals from existing communication systems. 
To the end, we design and implement a multi-antenna Doppler radar that does not need dedicated spectrum but leverages the over-the-air ambient RF signals for downstream sensing application tasks. 
\end{comment}

\subsection{Hardware Design}

As shown in Fig.~\ref{fig:ambientRFsensing}, the proposed sensing device comprises two parts: 
(i) amplify-and-forward circuit, and (ii) self-mixing circuit. 

\begin{figure}
    \centering
    \includegraphics[width=1\linewidth]{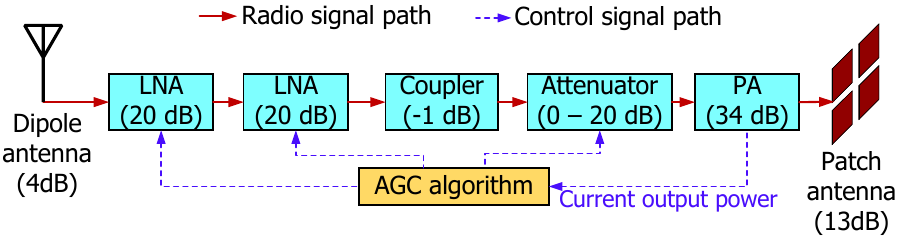}
    \caption{Diagram of amplify-and-forward RF circuit.}
    \label{fig:af_circuit}
\end{figure}

\textbf{Amplify-and-Forward Circuit:} 
Fig.~\ref{fig:af_circuit} shows our proposed circuit.
It uses an omnidirectional dipole antenna for signal reception, multi-stage LNA/PA with automatic gain control (AGC), and a directional patch antenna for efficient signal transmission. AGC, enabled by the PA's output power pin (e.g., ``DET"), prevents nonlinearity from excessive gain. Designed to forward signals at 10 dBm for sensing, the circuit operates effectively for ambient RF power from –70 to –7 dBm.
%, covering typical indoor Wi-Fi and cellular signal ranges.

% The amplify-and-forward circuit comprises a dipole antenna for omnidirectional signal reception, multi-stage LNA and PA with automatic gain control, and a patch antenna for directional signal transmission. 
% We use omnidirectional dipole antenna for the receiver because we want the sensing device works in any directions. 
% We use the directional patch antenna to enhance radio illumination efficiency for sensing purposes, following the most radar devices. 
% The automatic gain control plays a critical role in the management of signal amplification. 
% Overly large gain will saturate the PA and introduce nonlinearity, thereby generating interference for communication. 
% Fortunately, most PA chips have an output pin (typically named ``DET''), providing a voltage signal that is proportional to the output power of the amplifier. This pin is leveraged for automatic gain control.
% In indoor scenarios,  typical Wi-Fi signals may range from -40 to -80 dBm, and typical cellular signal power (e.g., 5G) range from about -50 dBm to -100 dBm. 
% Targeting a forwarding power of 10 dBm for object illumination, we propose an amplify-and-forward circuit as shown in Fig.~\ref{fig:af_circuit}. 
% Targeting a maximum output power 7dBm from the PA, this design can work when the ambient RF signal power is ranging -70~dBm to -7 dBm.

\begin{comment}
\begin{figure}
    \centering
    \includegraphics[width=0.75\linewidth]{figures/antenna_s21.pdf}
    \caption{S21 parameter of our amplify-and-forward circuit.}
    \label{fig:antenna_s21}
\end{figure}

% \textbf{Coupling of Tx and Rx Antennas:}
The coupling of the RX dipole antenna and the TX patch antenna, combining with the forward path with LNAs and PA, form a closed-loop feedback in the circuit. 
Care must be taken to minimize the coupling of the RX and TX antennas, preventing positive feedback in the closed-loop path. 
Many factors affect the coupling gain of the two antennas, including their position, orientation, and rotation. 
Fig.~\ref{fig:ambientRFsensing} shows our optimized geometry for installation. 
Fig.~\ref{fig:antenna_s21} shows our measured S21 parameter, where port 1 is dipole antenna's SMA port and port 2 is the patch antenna's SMA port. 
It shows that, at the intended frequency band 2.35 GHz, the coupling gain is less than -70~dB. 
Given that the total gain of forwarding path is less than 70~dB, this confirms that the closed-loop gain is less than 0 dB, preventing self amplification in the circuits. 
\end{comment}

\textbf{Self-Mixing Circuit:}
As shown in Fig.~\ref{fig:ambientRFsensing}, $M$ patch antennas are used to receive the signal reflected back from the nearby objects. 
These patch antennas go through an RF switch to share a single RF chain for circuit simplification. 
The reflected signal is then mixed with a copy of amplified ambient RF signal for down conversion, generating the baseband signal for feature extraction.
The RF switch is controlled by an FPGA device to meet the timing requirements, generating $M$ independent channels for object direction estimation.

The environment presents radio waves on both intended and unintended spectrum bands from diverse communication systems. 
Interference mitigation is challenging. 
We propose a joint hardware and algorithmic design for interference mitigation.
On the hardware side, we optimize the patch antenna design to minimize its out-of-band gain. 
Patch antennas have a small bandwidth by structure. 
This is widely considered its drawback in many other applications, but appears to be advantageous for this application. 
We optimize the antenna design to align its center frequency with the spectrum communication frequency channel, minimizing its bandwidth to suppress the out-of-band interference.

\subsection{Baseband Signal Analysis}

\begin{comment}

Developing a precise model that characterizes \pname's baseband signal (i.e., the output signal of its RF mixer) is challenging due to the complex propagation of the signals received by both the dipole and patch antennas. 
First, the radio signal received by the dipole is the communication OFDM signal experienced multi-path propagation. 
Additionally, the received signal may not be stable and be affected by the moving objects. 
The multi-path effect makes it challenging to analyze \pname's output baseband signal. 
Second, the objects are illuminated by both the amplified radio signal transmitted by the ``TX'' patch antenna and the ambient radio signal from the communication system.
This adds the complexity in the signal analysis.
Third, the sensing patch antenna receives both signals reflected by the objects and the direct signal from communication device. 
Fourth, the
\end{comment}

\textbf{Mathematical Modeling:}
Both Wi-Fi and 4/5G systems use OFDM modulation for signal transmission. 
Consider one OFDM symbol from the communication system. 
Denote $s(t)$ as the radio signal received by the dipole antenna (see Fig.~\ref{fig:ambientRFsensing}) that corresponds to one OFDM symbol. 
Then, we have 
% \begin{equation}
% S(t) = \sum_{l} \sum_{k} \alpha_l X_k e^{j2\pi (f_c + k f_\Delta) (t-\tau_l)}
% \end{equation}
% $s(t) = \sum_{k \in \mathcal{K}} X_k e^{j2\pi (f_c + k f_\Delta) t}$,
\begin{equation}
s(t) = \sum_{k \in \mathcal{K}} X_k e^{j2\pi (f_c + k f_\Delta) t}, 
\label{eq:ofdm}
\end{equation}
where $X_k \in \mathbb{C}$ is the QAM symbol, $k$ is OFDM subcarrier index and $\mathcal{K}$ is the set of valid subcarriers, $f_c$ is the carrier/center frequency of the radio signal, $f_\Delta$ is the subcarrier spacing.
%, $T_{ofdm}$ is the time duration of an OFDM symbol. 

The received signal $s(t)$ from the dipole antenna is amplified and forwarded to the ``TX'' antenna for transmission. 
The transmitted signal can be written as 
$\beta s(t)$, where $\beta \in \mathbb{C}$ represents the amplification and phase shift introduced by the amplify-and-forward circuit. 

The objects in the proximity are illuminated by two signal sources: \pname's TX antenna and the ambient RF signal (from the original communication system). 
Since the amplified signal is much stronger than the ambient signal itself, we ignore the ambient RF signal for illumination. 
Therefore, by denoting $r(t)$ as the received signal from the ``RX'' antenna, we have 
\begin{equation}
    r(t) = \beta \sum_{l \in \mathcal{L}}  \alpha_l \cdot s(t-\tau_l), 
\end{equation}
where $\mathcal{L}$ is the set of multi-path reflectors, $\alpha_l$ and $\tau_l$ are the attenuation coefficient and time delay of path $l$.

The received signal is mixed with the amplified ambient RF signal, generating baseband signal for feature extraction.
The baseband signal, i.e., the output of the RF mixer, can be written as:
\begin{equation}
y(t)  = r(t) \cdot s(t)^*  =|\beta|^2 \sum_{l \in \mathcal{L}}  \alpha_l \cdot s(t-\tau_l) \cdot s(t)^*,
\end{equation}
where $(\cdot)^*$ denotes conjugate operator.

Plugging Eqn~\eqref{eq:ofdm} into the above equation, we can obtain $y(t)$ as follows:
% shown in Eqn~\eqref{eq:y_t}.
\begin{align}
y(t)  
% &
% = |\beta|^2 \sum_{l \in \mathcal{L}}  \alpha_l \cdot 
% \left[\sum_{k \in \mathcal{K}} X_k e^{j2\pi (f_c + k f_\Delta) (t-t_l)} \right] 
% \left[\sum_{k' \in \mathcal{K}} X_{k'} e^{j2\pi (f_c + k' f_\Delta) t} \right]^*
% \nonumber \\
&
= |\beta|^2 \sum_{l \in \mathcal{L}}  \alpha_l \cdot 
\underbrace{
\left[\sum_{k \in \mathcal{K}} |X_k|^2 e^{-j2\pi (f_c + k f_\Delta) \tau_l} \right]
}_{\mbox{Part-A}} +
\nonumber \\
&
% \mbox{\ \ \ \ \ } 
\!\!\!\!\!
|\beta|^2 \sum_{l \in \mathcal{L}}  \!\!\! \alpha_l \!\!
\underbrace{
\left[\sum_{k \in \mathcal{K}} X_k e^{-j2\pi (f_c + k f_\Delta) \tau_l} \!\!\right]
}_{\mbox{Part-B}} 
% \nonumber \\
% &
\underbrace{\!\!\!
\left[\sum_{k' \in \mathcal{K}}^{k \neq k'} X_{k'}^* e^{j2\pi (k - k') f_\Delta t} \right]^*
\!\!\!\!\!}_{\mbox{Part-C}}
.
\label{eq:y_t}
% \nonumber \\
\end{align}

Denote $\mathcal{F}(\cdot)$ as the frequency of a signal. 
Then, we have 
$\mathcal{F}(\mbox{Part-A}) = 0$, 
$\mathcal{F}(\mbox{Part-B}) = 0$, 
and
$\mathcal{F}(\mbox{Part-C}) \ge f_\Delta$.

We apply a low-pass filter to $y(t)$ and denote the output signal as $z(t)$, which can be written as:
\begin{equation}
z(t) = \lfloor y(t) \rfloor_{f_\Delta/2} 
= |\beta|^2 \sum_{l \in \mathcal{L}}  \alpha_l \cdot \left[\sum_{k \in \mathcal{K}} |X_k|^2 e^{-j2\pi (f_c + k f_\Delta) \tau_l} \right],
\end{equation}
where 
$\lfloor \cdot \rfloor_{B}$ denotes a low pass filter with cutoff frequency $B$.

Denote $z(t) = \sum_{l \in \mathcal{L}} z_l(t)$, where $z_l(t)$ is contributed by the $l$-th object. 
Then, we have 
\begin{equation}
z_l(t)
= |\beta|^2  \alpha_l \cdot \sum_{k \in \mathcal{K}} |X_k|^2 e^{-j2\pi (f_c + k f_\Delta) \tau_l}.
\label{eq:zlt_b}
\end{equation}

Denote $d_l$ as the distance between \pname and the $l$-th object. 
We have $\tau_l = d_l/c$, where $c$ is the light speed.
Consequently, we have 
\begin{equation}
z_l(t)
= |\beta|^2  \alpha_l \cdot \sum_{k \in \mathcal{K}} |X_k|^2 e^{-j2\pi (f_c + k f_\Delta) \frac{d_l}{c}}.
\label{eq:zlt}
\end{equation}

Eqn~\eqref{eq:zlt} characterizes the relationship between the measured baseband signal and the object distance.
Particularly, we are interested in the relationship between $\angle{z_l}$ and the object distance $d_l$. 
In communication systems, OFDM subcarrier spacing is much smaller than the carrier frequency, i.e., $f_\Delta \ll f_c$.
Therefore, we approximate Eqn~\eqref{eq:zlt} by:
\begin{equation}
z_l(t)
\approx 
\left[
|\beta|^2  \alpha_l \sum_{k \in \mathcal{K}} |X_k|^2 
\right]
e^{-j2\pi f_c  \frac{d_l}{c}}.
\label{eq:zlt2}
\end{equation}

\begin{figure}
    \centering
    \includegraphics[width=0.7\linewidth]{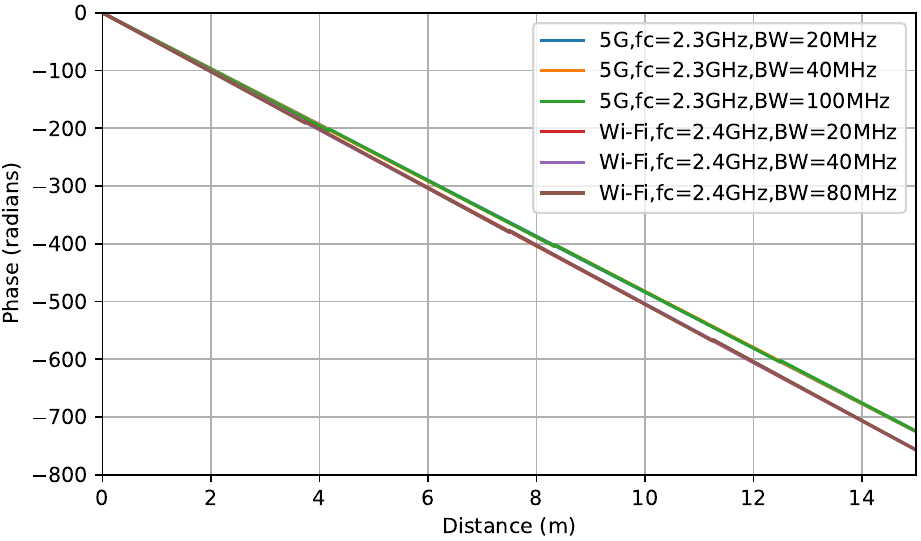}
    \caption{Relationship between an object-moving distance and the signal phase rotation. 
    % Relation between an object distance and the phase of observed signal from the numerical analysis of 5G and Wi-Fi systems. $f_c$ is the carrier frequency and BW is the signal bandwidth.
    }
    \label{fig:d_vs_phase}
\end{figure}

Eqn~\eqref{eq:zlt2} indicates the linear relationship between $d_l$ (object distance) and $\angle z_l(t)$.
Based on Eqn~\eqref{eq:zlt2}, the instantaneous velocity of the $l$-th object, denoted as $v_l(t)$, can be estimated by:
\begin{equation}
    v_l(t) = \frac{unwrap(\angle z_l(t)) - unwrap(\angle z_l(t-T_\Delta))}{T_\Delta},
    \label{eq:velocity}
\end{equation}
where $unwrap(\cdot)$ is a function used to fix phase discontinuities in angle data. 
$T_\Delta$ is a small time step.

Eqn~\eqref{eq:velocity} shows the linear relationship between an object's moving distance and the phase change, and thus lays the foundation for the design of our Doppler signature extraction. 

% \begin{lemma}
%     \pname is capable of estimating the Doppler features of objects.
%     \label{lemma:1}
% \end{lemma}

\textbf{Numerical Results:}
The derivation from Eqn~\eqref{eq:zlt} to Eqn~\eqref{eq:zlt2} is based on the assumption that $f_\Delta$ is negligible compared to $f_c$. 
In 5G and Wi-Fi communication systems, despite the fact that $f_\Delta \ll f_c$, $f_\Delta$ is not negligible.
In Wi-Fi systems, $f_\Delta = 312.5$ kHz while $f_c = 2.4$ GHz or $f_c = 5$ GHz.
In 4G/5G systems, $f_\Delta = 15$ kHz or $f_\Delta = 30$ kHz while its $f_c$ is typically larger than 1.8 GHz. 
We thus conduct numerical studies to evaluate the relation between $\angle z_l(t)$ and $d_l$ in Eqn~\eqref{eq:zlt2}. 
Fig.~\ref{fig:d_vs_phase} presents our numerical results.
It confirms the approximately linear relation between the angle of $z_l(t)$ and object moving distance $d_l$.

\section{Data Processing}
\label{sec:preproc}

The above analysis reveals the Doppler signatures of moving objects in the observed signal.
However, extracting meaningful features for human activity detection presents several challenges that were not accounted for in the prior analysis:
(i) the signal received by \pname's RX0 may undergo multipath propagation,
(ii) the ambient 5G signal is inherently noisy,
(iii) the use of adaptive modulation and coding schemes (MCS) in 5G introduces variability, and
(iv) interference from adjacent systems such as Wi-Fi can distort the signal.
In this section, we first introduce a heuristic algorithm for signal sanitization to mitigate these challenges, followed by a differential beamforming scheme for generating heatmap images.

\begin{figure}
    \centering
    \includegraphics[width=0.35\linewidth]{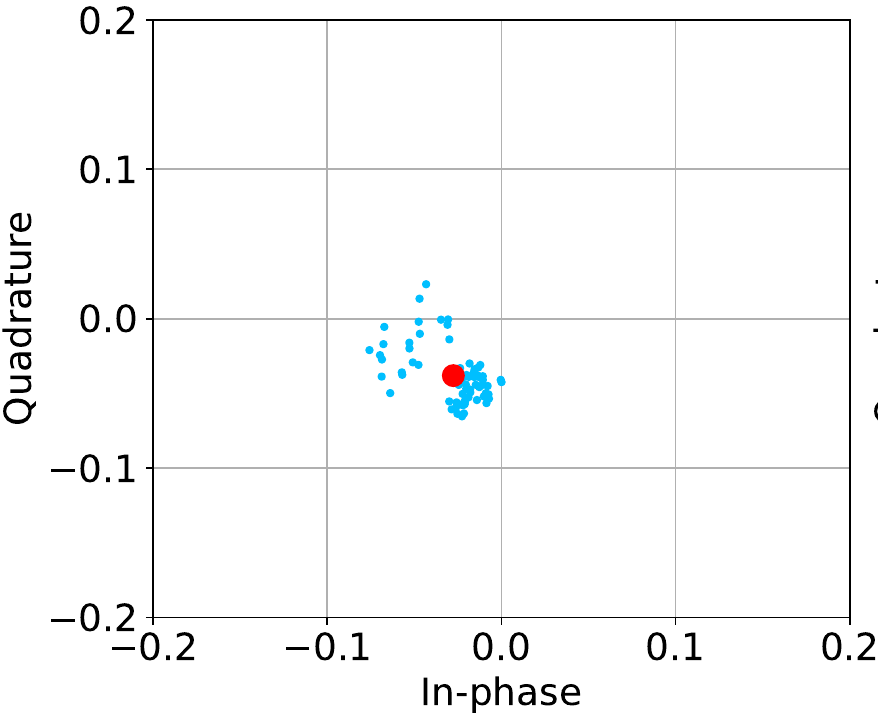}
    \hspace*{0.25in}
    \includegraphics[width=0.35\linewidth]{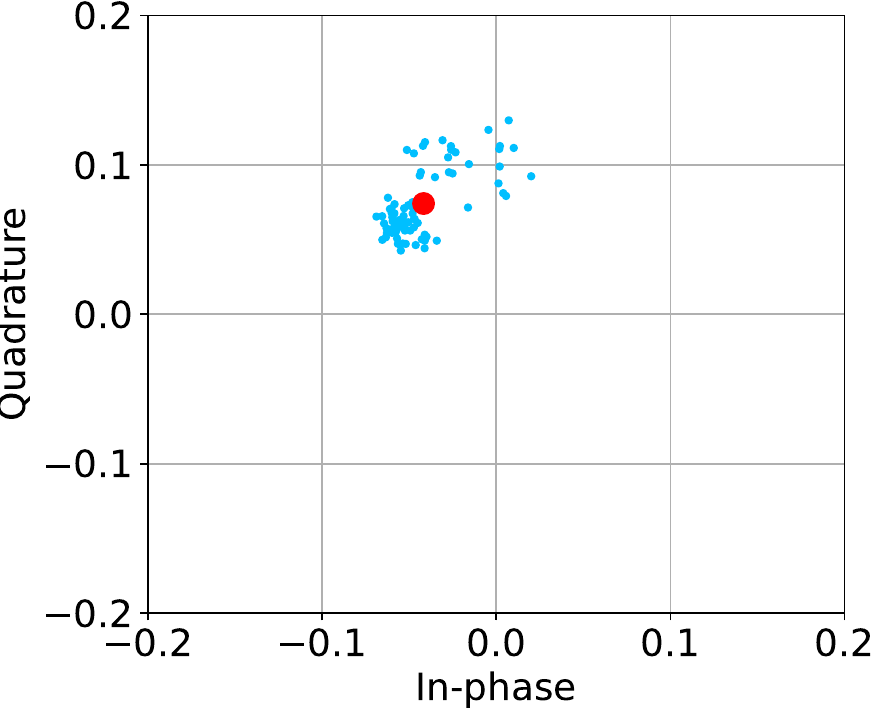}
    \caption{Signal constellation before (blue dot scatters) and after (a single red point) sanitization.}
    \label{fig:constellationData}
\end{figure}

\subsection{Signal Sanitization}

To ensure reliable human activity detection from ambient 5G signals, we design a multi-stage signal processing pipeline that enhances signal quality by addressing noise, bias, and outliers.
Our approach is based on the fact that the signal sampling rate is 2 MSPS, which is much higher than what is required for human activity detection.
This high sampling rate introduces redundancy that can be exploited for effective signal sanitization.

\textbf{Noise Suppression and Bias Correction.}
The blue dot scatters in Fig.~\ref{fig:constellationData} shows the constellations of the raw baseband signal samples from the RF hardware (i.e., the output of self-mixer). 
Depending upon the ambient 5G signal strength and other factors, the raw signal samples may appear to wide-spread or well-concentratrated. 
To mitigate these issues, we first apply a Butterworth low-pass filter, which attenuates high-frequency noise while preserving relevant signal features. 
Next, we correct sample bias by identifying stationary signal components clustered near a common offset. 
We use the k-means clustering algorithm to group constellation points into three clusters and identify the one closest to the origin. This cluster is presumed to represent static signal components. All constellation points are then translated to re-center this cluster at the origin, thereby normalizing the spatial bias.

\textbf{Outlier Detection and Removal.}
Despite initial noise reduction, residual outliers may still present, primarily due to interference from radio systems on adjacent spectrum bands. These outliers can distort downstream learning models and obscure motion-related patterns. We implement a two-step strategy for outlier removal:
(i) all points within a defined radius around the origin  are discarded, and 
(ii) we employ the Local Outlier Factor (LOF) algorithm to eliminate spatially sparse and anomalous points. 
LOF compares the local density of a point to its neighbors using k-nearest neighbor metrics, and flags points with significantly lower density as outliers. This technique effectively preserves regions of meaningful, homogeneous signal activity while suppressing noise-induced anomalies.

\textbf{Sample Projection.}
After denoising and outlier removal, the data often still contains some scattered points. For efficient heatmap generation and model input simplification, we reduce the constellation diagram to a single representative projection. We apply k-means clustering to all remaining points and retain only the centroid of the dominant cluster. This yields a compact, informative representation of the signal dynamics associated with each time frame.
The single red point in Fig.~\ref{fig:constellationData} represents the output of the sanitization.

\begin{figure}
    \centering
    \includegraphics[width=1\linewidth]{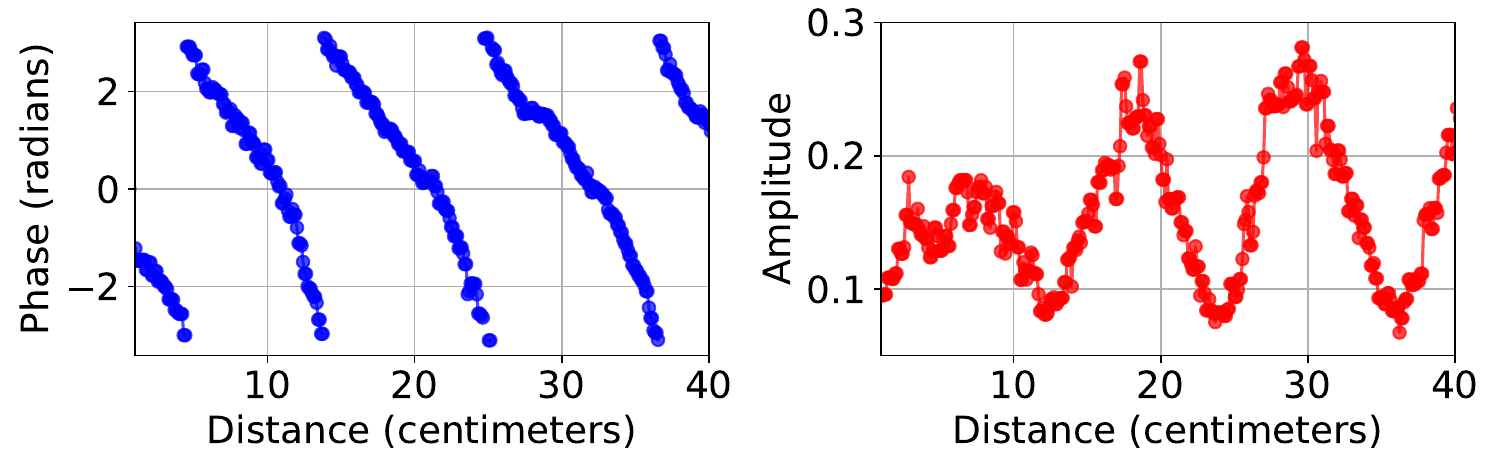}
    \caption{Observed signal phase and amplitude when a person moves toward the detector at a roughly constant speed.}
    \label{fig:amplitude_phase}
\end{figure}
\textbf{Experimental Validation.}
To validate the relationship between movement distance and phase rotation, we conducted experiments using our testbed (see Fig.~\ref{fig:testbed} in \S\ref{subsec:impl}).
We measured the signal amplitude and phase as a person moved toward the antennas at a constant speed.
The carrier frequency of the 5G signal was 2.35 GHz.
Fig.~\ref{fig:amplitude_phase} shows the measurement results.
Notably, the phase exhibits a clear linear relationship with distance, confirming coherent phase detection and the presence of Doppler features.
As expected, the signal amplitude also change with the person's movement. 

% validates our implementation: moving at a constant speed over a distance of 40 centimeters, we observe that the phase and amplitude variations of our representative projection in Fig.~\ref{fig:constellationData} are close to the theoretical variations for a frequency of 2.3 GHz and therefore a wavelength of approximately 13 cm.

\subsection{Differential Beamforming}
\label{sec:spatial_resolution}

\begin{figure}[t]
\centering

% Left: Single image
\begin{minipage}[b]{0.35\linewidth}
    \centering
    \includegraphics[width=\linewidth]{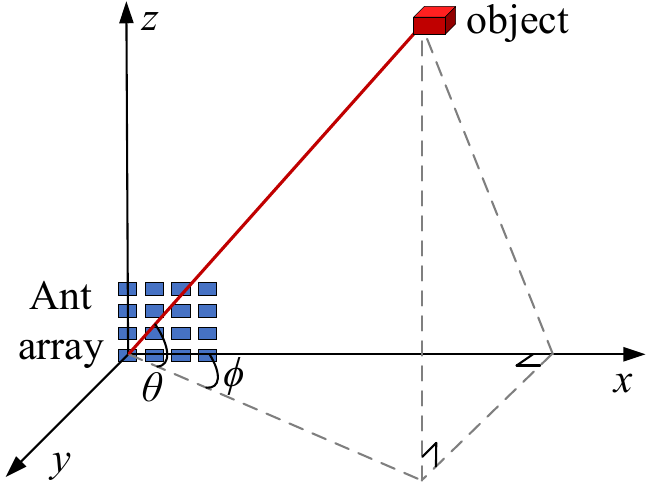}
    \caption{Illustration of spatial beamforming.}
    \label{fig:ant_imaging}
\end{minipage}
\hfill
% Right: Two subfigures
\begin{minipage}[b]{0.6\linewidth}
    \centering
    \begin{subfigure}{0.48\linewidth}
        \centering
        \includegraphics[width=\linewidth]{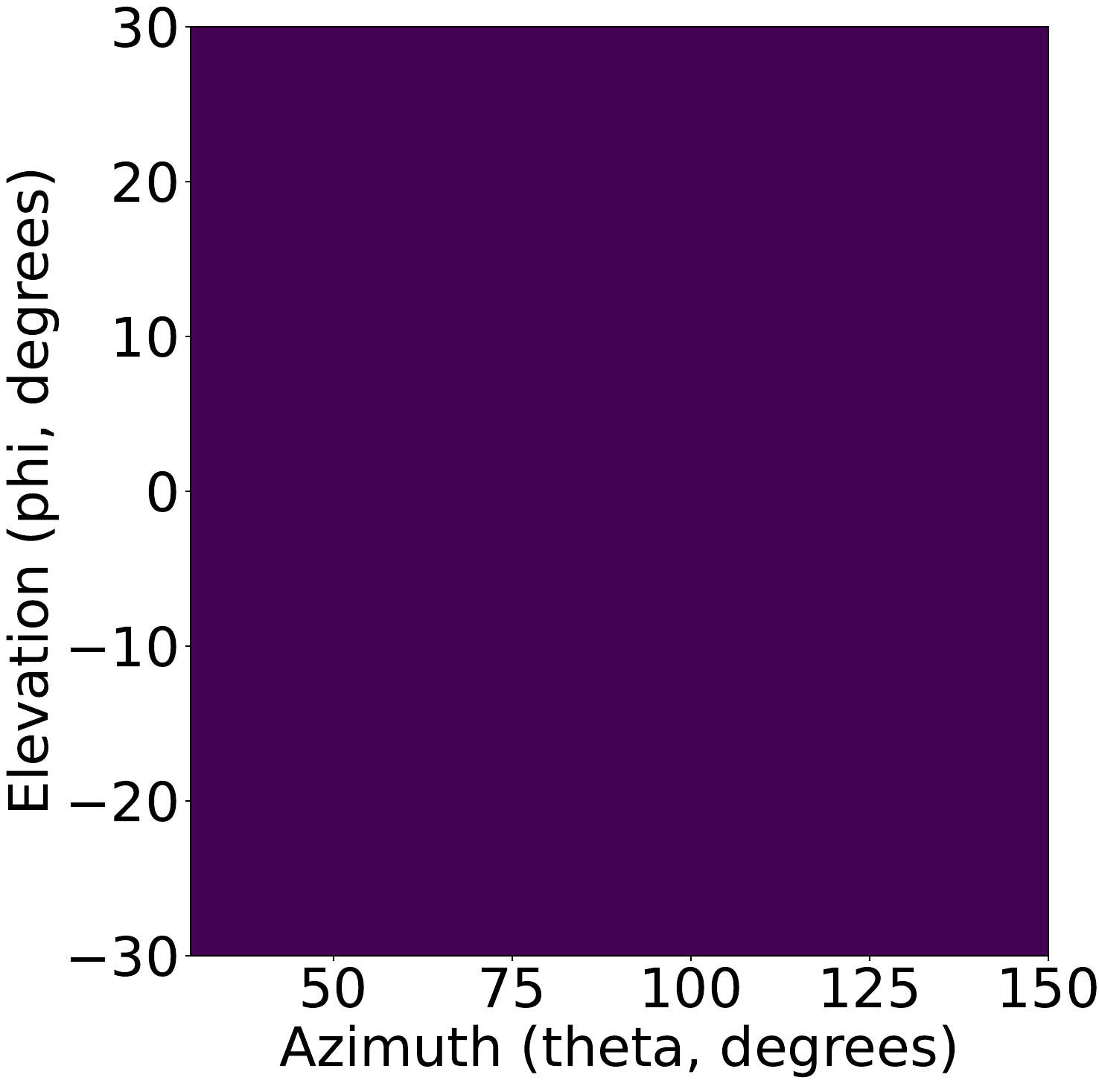}
        % \caption{W/o moving object.}
    \end{subfigure}
    \begin{subfigure}{0.48\linewidth}
        \centering
        \includegraphics[width=\linewidth]{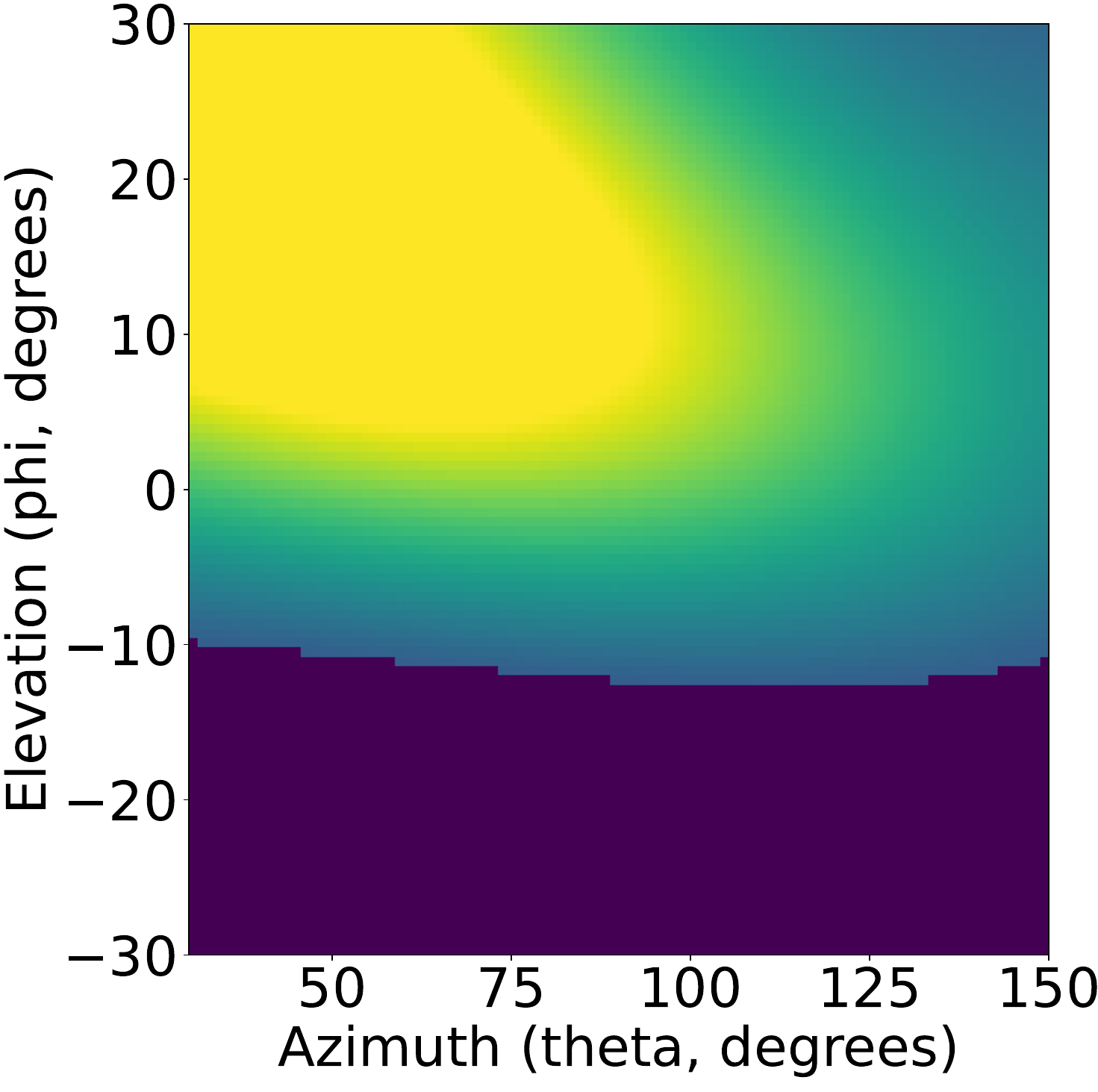}
        % \caption{W/ moving object.}
    \end{subfigure}
    \caption{Heatmap from object-static (left) and object-moving (right) cases.}
    \label{fig:heatmap}
\end{minipage}

\end{figure}

Here, beamforming refers to the use of a two-dimensional antenna array to estimate the direction of incoming signals, namely, the azimuth angle $\theta$ and elevation angle $\phi$ as illustrated in Fig.~\ref{fig:ant_imaging}.
This technique, along with its variants, is widely used in radar imaging to infer the spatial structure of a scene.
In our work, we focus on detecting motion rather than reconstructing static scenes, and therefore refer to our approach as differential beamforming.

Consider the antenna array as shown in Fig.~\ref{fig:ant_imaging}. 
One antenna is for transmitting and the other antennas for receiving. 
Denote the coordinate of the TX antenna as the origin,
i.e.,
$\vec{p}_{TX} = (0, 0, 0)$.
Denote $\mathcal{R}$ as the set of RX antenna array, and $\vec{p}_i$ as the coordinate of RX antenna $i \in \mathcal{R}$.

Consider a small object or a small part of a big object characterized by azimuth angle $\theta$ and elevation angle $\phi$ as shown in Fig.~\ref{fig:ant_imaging}.
To estimate the signal strength from this direction, we define a unit direction vector by letting:
\begin{equation}
\vec{u}(\theta, \phi) = \begin{pmatrix}
    \cos(\theta)\cos(\phi), \ \ 
    \cos(\theta)\sin(\phi), \ \ 
    \sin(\phi)
\end{pmatrix}.
\end{equation}

Based on this unit vector, the expected phase offset of the signal received by RX antenna $i \in \mathcal{R}$ can be written  as:
\begin{equation}
w_i(\theta, \phi) = \exp\left(-j \frac{2\pi}{\lambda} \langle \vec{u}(\theta, \phi), \ \vec{p}_i \rangle \right),
\end{equation}
where $\lambda$ is the wavelength of 5G/Wi-Fi carrier frequency, and $\langle\cdot, \cdot\rangle$ is the inner product.

Recall that $z_i(t)$ represents the output baseband signal from Rx antenna $i \in \mathcal{R}$ at time $t$. Then, the differential beamforming operation can be written as:
\begin{equation}
I(\theta, \phi, t) = \sum_{i \in \mathcal{R}} w_i(\theta, \phi) \cdot \left[z_i(t) - z_i(t-T_\Delta)\right]^*,
\label{eq:intensity}
\end{equation}
where $T_\Delta$ is a fixed time gap, which is empirically chosen.
We note that, by taking the difference of $z(t)$ over a time period $T_\Delta$, we suppress static components and concentrate on moving objects such as human, thereby enhancing motion detection sensitivity.

Referring to Fig.~\ref{fig:ant_imaging}, we use the amplitude of $I(\theta, \phi, t)$ to estimate the strength of incoming signal from the direction $(\theta, \phi)$ at time $t$. 
By querying $I(\theta, \phi, t)$ with all possible $\theta$ and $\phi$ values, we obtain the heatmap image of the moving objects at time $t$, 
i.e., 
\begin{equation}
    \mathbf{X}(t) = \Big[ \left|I(\theta, \phi, t)  \right|: \forall \theta, \forall \phi  \Big],
    \label{eq:heatmap}
\end{equation}
where $\mathbf{X}$ is the heatmap of moving object at time $t$. 

Fig.~\ref{fig:heatmap} presents two examples of heatmaps generated using Eqn~\eqref{eq:heatmap}.
The data was collected using a radio setup 
consisting of one TX antenna and eight RX antennas (see Fig.~\ref{fig:testbed} in \S\ref{subsec:impl}).
The system operates on an ambient 5G signal with a carrier frequency of 2.35 GHz and a bandwidth of 40 MHz.
Fig.~\ref{fig:heatmap}(a) shows the heatmap generated in a static environment with no movement, while Fig.~\ref{fig:heatmap}(b) shows the heatmap when a person is moving on the right side of the sensing area.
As shown, the heatmap is able to capture movement at a coarse level.
In the next section, we introduce a DNN model to refine this coarse representation for more accurate activity detection.

\section{Human Activity Recognition}
\label{sec:dnn}

% In this section, we use two downstream tasks, human skeleton and body mask estimation, as the examples to demonstrate the potential of \pname in human activity detection. 

In this section, we evaluate the capabilities of \pname through two challenging downstream tasks: human skeleton estimation and body mask segmentation. These tasks serve as a benchmark to demonstrate the potential of leveraging ambient 5G signals for fine-grained human activity detection.

% To evaluate the capability of our proposed sensing system, \pname focuses on two challenging tasks: mask segmentation and keypoint estimation.

\subsection{Overview}

% RF-based human activity recognition leverages the property that RF signals reflect off the human body to identify various human activities 

% While there is a considerable body of work on radio-based human activity detection \cite{wang2019can, wang2019person, yang2022metafi, zhou2023adapose, zhao2018through, zhao2018rf, zhao2019through}, our work differs from them...

Radio signal features are inherently sparse, noisy, and sensitive to environmental dynamics, which presents significant challenges for accurately perceiving fine-grained human activity. To address this, we employ a powerful deep neural network (DNN) capable of learning complex representations from radio features. Our approach centers on a cross-modal supervised learning framework that distills high-fidelity knowledge from the vision domain into the radio domain. During the training phase, we equip the \pname system with a co-located video camera to provide precise ground-truth labels, such as human skeletons and body masks, which supervise the learning of the radio-based model.

% Radio signal features are often sparse, noisy, and sensitive to environmental variations, making it challenging to accurately detect fine-grained human activity information.
% This necessitates a powerful deep neural network (DNN) capable of learning complex representations for human pose and activity recognition from radio signal features.
% To overcome this challenge, we adopt a cross-modal supervised learning approach that distills knowledge from the vision domain into the radio domain.
% Specifically, during training, we equip \pname with a video camera to provide ground-truth labels (e.g., human skeletons and body masks) used to supervise the learning of the DNN model for radio-based activity recognition.

\begin{figure}
    \centering
    \includegraphics[width=\linewidth]{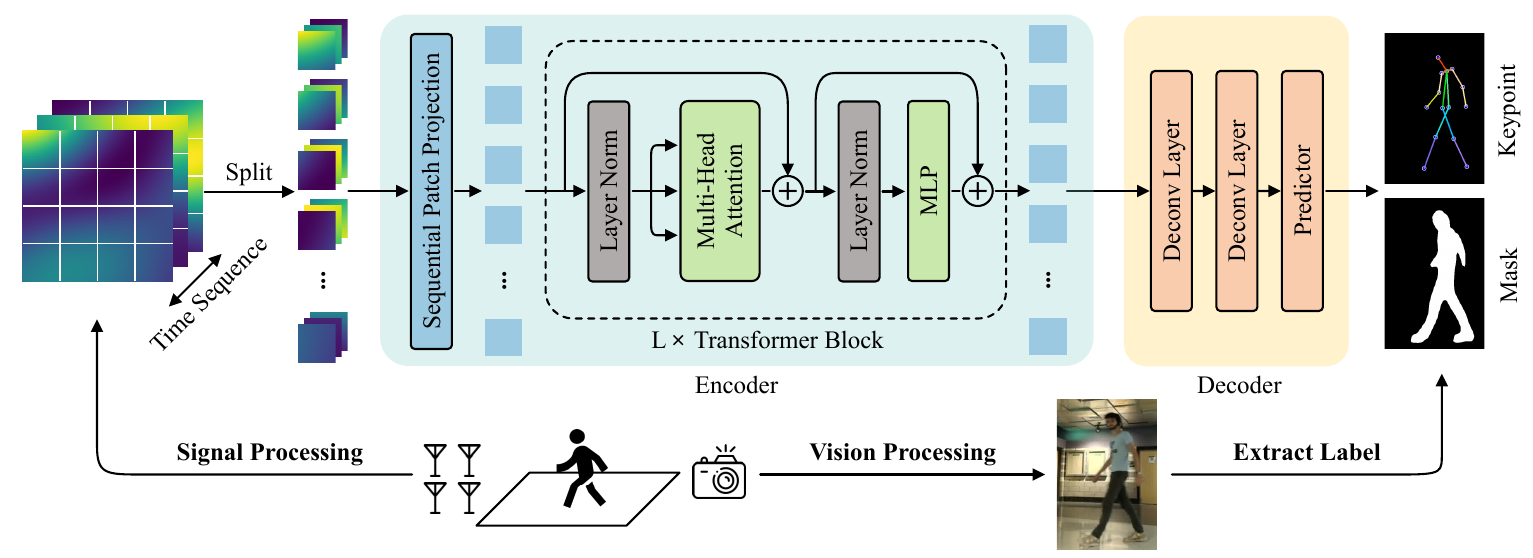}
    \caption{Deep neural network framework of \pname. 
    % The framework employs a cross-modal training strategy with two main pipelines: \textit{vision processing} and \textit{signal processing}. During training, the vision pipeline extracts ground truth labels from video frames to supervise the signal pipeline, which learns to map RF signals to the corresponding mask and keypoint heatmaps. At inference time, only the signal processing pipeline is needed.
    }
    \label{fig:soft_overview}
\end{figure}

The proposed architectural framework, illustrated in Fig.~\ref{fig:soft_overview}, consists of two primary pipelines: signal processing and vision processing. The signal processing pipeline features a Split module, an Encoder, and a Decoder; it processes a sequence of radio heatmap frames (as defined in Eqn~\eqref{eq:heatmap}) to output the estimated human skeletal structure and body mask. Simultaneously, the vision pipeline leverages off-the-shelf computer vision algorithms to extract ground-truth annotations from synchronized video frames. Crucially, once the model is trained, the video camera is discarded. In the inference stage, only ambient radio heatmap features are required to perform accurate and privacy-preserving human activity estimation.

% Fig.~\ref{fig:soft_overview} illustrates our proposed architectural framework, which consists of two primary pipelines:
% \textit{signal processing} 
% and
% \textit{vision processing}.
% The signal processing pipeline comprises a Split module, an Encoder, and a Decoder.
% It takes as input a sequence of radio heatmap frames---each defined in Eqn~\eqref{eq:heatmap}---and outputs the estimated human skeletal structure and body mask.
% During training, the vision pipeline leverages off-the-shelf computer vision algorithms to extract ground-truth human skeletons and body masks from video frames. These serve as supervisory ground-truth labels to train the the models in the signal processing pipeline.
% In the inference stage, the video camera is no longer needed.
% Only radio heatmap features are fed into the trained model for human activity estimation.

\subsection{Structure Framework}

In practice, RF signal reflections from specific human body parts may not be captured by the receivers consistently, resulting in intermittent information loss.
To mitigate the impact of such temporal voids, we exploit the inherent temporal correlation within sequences of radio frames.
Given the proven effectiveness of self-attention in capturing long-range dependencies, we leverage it to jointly model the spatio-temporal relationships within the radio frame sequence. Specifically, we propose a sequential patch projection method to structure the input data for this joint analysis. 
Given a sequence of RF frames \( \textbf{X} \in \mathbb{R}^{F \times H \times W} \), where \( (H, W) \) denotes the spatial dimension of each radio heatmap specified in Eqn~\eqref{eq:heatmap} and \( F \) is the number of frames, we first partition the data into \(N = HW / P^2\) non-overlapping spatial patches:
\begin{align}
\text{Split}(\mathbf{X}) = (\mathbf{X}_P^1, \mathbf{X}_P^2, \ldots, \mathbf{X}_P^N) : \mathbb{R}^{F \times H \times W} \mapsto \mathbb{R}^{F \cdot P^2},
\end{align}
where \((P, P)\) is the patch size, and \( N \) is the total number of patches. The data is then reshaped into a new sequence \( \textbf{X}_P \), a step that ensures each element encapsulates the complete temporal trajectory for a single spatial location. 
\begin{align}
\textbf{X}_P = [\mathbf{X}_P^1, \mathbf{X}_P^2, \cdots, \mathbf{X}_P^N] : \mathbb{R}^{N \times ( F \cdot P^2 ) }.
\end{align}

These spatio-temporal patches are subsequently flattened and linearly projected into \( C \)-dimensional embeddings via a trainable layer \(f_\theta (\cdot)\), yielding the initial representation $\mathbf{Z}_0$:
\begin{align}
\mathbf{Z}_0 = f_\theta (\mathbf{X}_P) : \mathbb{R}^{N \times ( F \cdot P^2 )} \mapsto \mathbb{R}^{N \times C}.
\end{align}

This sequence is then processed by a series of $L$ transformer blocks, each comprising a multi-head self-attention (MHSA) layer and a multi-layer perception (MLP). To ensure stable training, layer normalization is applied before each layer, followed by residual connections. The computation for the \(l\)-th transformer block is as follows:
\begin{align}
\mathbf{Z}'_{l} 
&= \mathbf{Z}_{l - 1} + \text{MHSA}(\text{LN}(\mathbf{Z}_{l - 1})), & l = 1, \ldots, L. \\
\mathbf{Z}_{l} 
&= \mathbf{Z}'_{l} + \text{MLP}(\text{LN}(\mathbf{Z}'_{l})), & l = 1, \ldots, L.
\end{align}
where \(\mathbf{Z}_{l}\) represents the output of the \(l\)-th transformer block. We denote the output of the final transformer block as \(\mathbf{Z}_{\text{out}} \in \mathbb{R}^{N \times C}\), serves as the representation for downstream tasks.

To perform dense prediction from these extracted features, we design a universal decoder architecture comprising consecutive deconvolution layers. We first unflatten the output embedding sequence \(\mathbf{Z}_{\text{out}}\) back into a feature map \(\mathbf{F}_0 \in \mathbb{R}^{C \times \frac{H}{P} \times \frac{W}{P}}\) to restore the original 2D spatial arrangement. This feature map is then progressively upsampled; each deconvolution layer doubles the spatial resolution while halving the number of channels. After $i$ such layers, the resulting feature map \( \mathbf{F}_i \) reaches dimensions of:
\begin{equation}
\mathbf{F}_i \in \mathbb{R}^{\frac{C}{2^i} \times \frac{2^i H}{P} \times \frac{2^i W}{P}}, \quad \text{for } i \in \{1, 2\}.
\end{equation}

Finally, a \(1 \times 1\) convolution layer projects the map to \(C_{\text{out}}\) channels, where the output depth is tailored to the requirements of the specific task.

\subsection{Loss Function}

\subsubsection{Mask Segmentation} 
We employ a composite loss $\mathcal{L}_{\text{mask}}$ that combines Binary Cross-Entropy (BCE) with Dice loss. While BCE handles pixel-wise classification \cite{he2017mask, long2015fully, ronneberger2015u}, the Dice component is incorporated to mitigate the severe class imbalance between the sparse human masks and the background \cite{milletari2016v}. The balance between these terms is controlled by hyperparameters \(\alpha_1\) and \(\alpha_2\), with $\epsilon$ ensuring numerical stability:
\begin{equation}
\begin{aligned}
\mathcal{L}_{\text{mask}} = & - \alpha_1 \sum_{i=1}^N \Big(y_i \log(x_i) + (1-y_i)\log(1-x_i)\Big) \\
& + \alpha_2 \left(1-\frac{2\sum_{i=1}^N x_iy_i + \epsilon}{\sum_{i=1}^N x_i^2 + \sum_{i=1}^N y_i^2 + \epsilon}\right),
\end{aligned}
\end{equation}
where \(x_i\) is the predicted probability of the \(i\)-th pixel belonging to the mask, and \(y_i\) is the corresponding ground truth label. \(N\) represents the total number of pixels.

\subsubsection{Keypoint Estimation}
For keypoint estimation, we utilize a two-level weighted Mean Squared Error (MSE) loss $\mathcal{L}_{\text{keypoint}}$:
\begin{equation}
\mathcal{L}_{\text{keypoint}} = \frac{1}{K} \sum^K_{k=1} w_k \cdot \text{mean} \left( (\mathbf{Y}_k + 1) \odot ( \hat{\mathbf{Y}}_k - \mathbf{Y}_k )^2 \right), 
\end{equation}
where \(K\) is the total number of keypoints, \(\hat{\mathbf{Y}}_k\) and \(\mathbf{Y}_k\) are the predicted and ground truth heatmaps for the \(k\)-th keypoint, respectively. %The \(\text{mean}(\cdot)\) operator calculates the average error over all pixels in a given heatmap, and \(\odot\) denotes element-wise multiplication. 
Unlike standard MSE, this approach prioritizes pixels in the immediate vicinity of joints and assigns joint-specific weights $w_k$ to emphasize critical skeletal structures.

% The keypoint estimation task utilizes a weighted Mean Squared Error (MSE) loss to supervise the regression of joint heatmaps.  Unlike a standard MSE that treats all pixels equally \cite{chen2018cascaded, fang2017rmpe, papandreou2017towards, xiao2018simple, xu2022vitpose, sun2019deep}, our two-level weighted approach emphasizes pixels in the immediate vicinity of target keypoints and assigns varying importance levels to different joints to prioritize more critical skeletal structures:
% \begin{equation}
% \mathcal{L}_{\text{keypoint}} = \frac{1}{K} \sum^K_{k=1} w_k \cdot \text{mean} \left( (\mathbf{Y}_k + 1) \odot ( \hat{\mathbf{Y}}_k - \mathbf{Y}_k )^2 \right), 
% \end{equation}
% where \(K\) is the total number of keypoints, \(\hat{\mathbf{Y}}_k\) and \(\mathbf{Y}_k\) are the predicted and ground truth heatmaps for the \(k\)-th keypoint, respectively, and \(w_k\) is a predefined scalar weight for the \(k\)-th keypoint, allowing us to prioritize more critical joints. The \(\text{mean}(\cdot)\) operator calculates the average error over all pixels in a given heatmap, and \(\odot\) denotes element-wise multiplication. 

To organize identity-free keypoints into individual instances, we adopt an associative embedding strategy \cite{law2018cornernet, newell2017associative, cheng2020higherhrnet} governed by the grouping loss $\mathcal{L}_{\text{group}}$:
\begin{equation}
\begin{aligned}
\mathcal{L}_{\text{group}} 
& =\lambda_{1} \frac{1}{|P|} \sum_{p \in P} \frac{1}{|K_p|} \sum_{k \in K_p} (e_{p,k} - \bar{e}_p)^2 \\
& + \lambda_{2} \frac{1}{|P|(|P|-1)} \sum_{p_i \in P} \sum_{p_j \in P, i \neq j} \max(0, \delta - |\bar{e}_{p_i} - \bar{e}_{p_j}|),
\end{aligned}
\end{equation}
where \(P\) is the set of persons, and \(K_p\) is the set of visible keypoints for person \(p \in P\). This loss enforces intra-instance cohesion by clustering joint tags around their mean $\bar{e}_p$, while maintaining inter-instance separation through a margin hyperparameter $\delta$. The objectives are balanced via \(\lambda_1\) and \(\lambda_2\).

\section{Experimental Evaluation}
\label{sec:exp}

% In this section, we conduct extensive experiments with the aim of answering the following questions.
% \begin{itemize}
% \item 
% \textbf{Q1 (\S\ref{subsec:performance}):}
% How does \pname perform when compared to the SOTA in human detection?

% \item 
% \textbf{Q2 (\S\ref{subsec:analysis}):}
% How does the parameters (e.g., signal duration and  distance) affect the performance of \pname?

% \item 
% \textbf{Q3 (\S\ref{subsec:zeroshot}):}
% How does \pname perform when compared to the SOTA in \textit{unseen} scenarios?
% \end{itemize}

\subsection{Implementation}
\label{subsec:impl}

\subsubsection{Hardware}
\begin{figure}
    \centering
    \includegraphics[width=\linewidth]{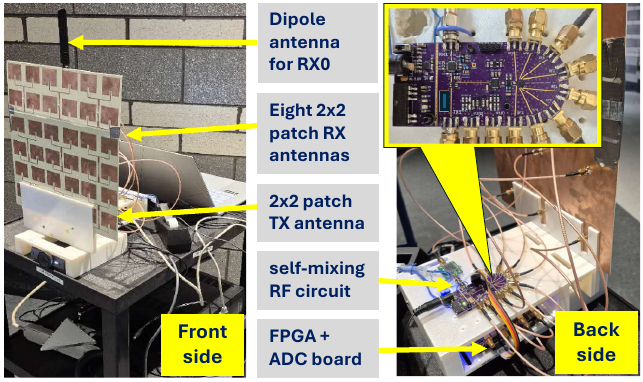}
    \caption{Our implementation of \pname.}
    \label{fig:testbed}
\end{figure}

We have developed a prototype of \pname, as illustrated in Fig.~\ref{fig:testbed}, to evaluate its performance in realistic scenarios.
The hardware consists of an RF circuit, a dipole antenna, a patch antenna array, and an FPGA board.
As shown in Fig.~\ref{fig:ambientRFsensing}, the RF board comprises two main components: an amplify-and-forward circuit and a self-mixing circuit.
The amplify-and-forward section is implemented using the MMZ25332 amplifier, which provides a 30 dB power gain over the 1.8--2.7 GHz frequency range.
A Mini-Circuits D17W+ signal coupler is used to derive the LO signal for the self-mixing circuit, which integrates a Skyworks SKY13418 for eight-channel RF switching, a Qorvo QPL9096 low-noise amplifier (LNA), and an Analog Devices LT5575 quadrature mixer.
The printed circuit board (PCB) was fabricated on an OSH Park FR408 substrate.
To digitize the baseband signals, we use the ECLYPSE Z7 FPGA board, which features two ADC channels operating at a sampling rate of 2 MSPS.
The patch antennas were simulated using HFSS and fabricated on Rogers RO4350B substrate.

\subsubsection{Ambient 5G Signals}
To avoid potential interference with commercial 5G services provided by AT\&T, Verizon, and T-Mobile, we set up a private 5G base station using commercial O-RAN equipment, including a Benetel O-RU, srsRAN O-DU and O-CU, and Open5GS core. Smartphones are connected to this 5G base station operating on spectrum band n40 (TDD, 2300--2400 MHz). This spectrum band was used under our FCC Experimental License with Call Sign \#WA3XEP. \pname leverages the radio signals from this private 5G base station for sensing applications.

% \begin{figure}
%     \centering
%     \includegraphics[width=1\linewidth]{figures/testbed.pdf}
%     \caption{Our prototype of \pname.}
%     \label{fig:testbed}
% \end{figure}

% \begin{figure*}
%     \centering
%     \begin{minipage}{0.395\linewidth}
%         \centering
%         \includegraphics[width=\linewidth,height=1.78in]{figures/testbed_v2.pdf}
%         \caption{Our implementation of \pname.}
%         \label{fig:testbed}
%     \end{minipage}\hfill
%     \begin{minipage}{0.6\linewidth}
%         \centering
%         \includegraphics[width=\textwidth]{picture/performance_h.pdf}
%         \caption{Qualitative results for keypoint estimation and mask segmentation.}
%         \label{fig:performance}
%     \end{minipage}
% \end{figure*}

\begin{figure*}
    \centering
    \includegraphics[width=\textwidth]{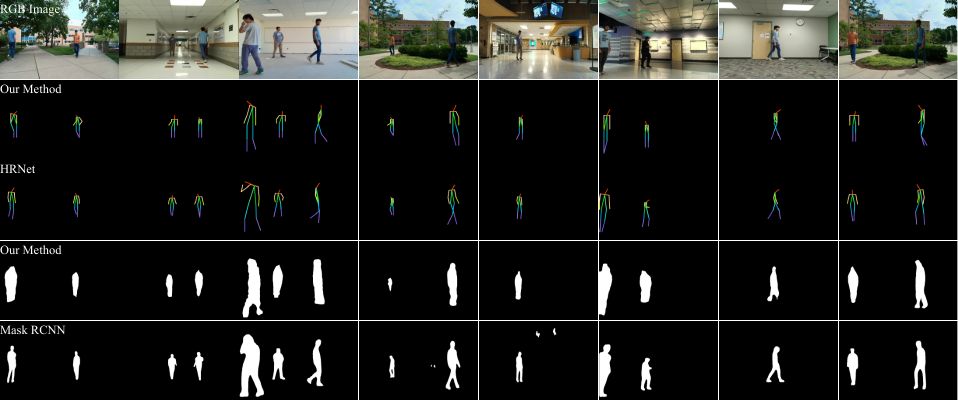}
    \caption{Qualitative results for keypoint estimation and mask segmentation.}
    \label{fig:performance}
\end{figure*}

\subsubsection{Software}
We implement our DNN framework in PyTorch, taking 21 RF frames of size $100 \times 100$ as input. These are divided into $10 \times 10$ patches, linearly embedded into a 256-dimensional space, and processed by a 2-layer Transformer encoder. The decoder output channels are set to 1 for binary segmentation and 26 for keypoint estimation. Training uses the Adam optimizer ($5\mathrm{e}\!-\!4$ LR) on four NVIDIA A6000 GPUs.

\subsection{Data and Annotations}

Our data collection testbed consists of four RF sensing device and a co-located web camera for capturing ground truth. We collected a comprehensive dataset across 8 diverse indoor and outdoor scenes on a university campus.
%, with representative environments shown in Fig.~\ref{fig:dataset}. 
During each collection session, the camera recorded time-synchronized video at 10 FPS. To ensure a rich variety of data, participants were instructed to move and perform arbitrary actions freely within the sensing area, imposing no restrictions on their poses or positions. We partitioned our dataset chronologically, allocating the first 80\% of the collected data for the training set and the remaining 20\% for the test set. This resulted in 206,734 training samples and 51,677 test samples, respectively.

To generate supervision signals for our cross-modal training, we automatically extracted annotations from the synchronized video frames. Specifically, we employed Mask R-CNN \cite{he2017mask} to obtain binary body segmentation masks and HRNet \cite{sun2019deep} to extract 2D human keypoints.

\begin{table*}
\scriptsize
\centering
\caption{Quantitative comparison of \pname with SOTA methods. 
% \pname achieves state-of-the-art performance on both mask segmentation and keypoint estimation tasks. All metrics are the higher the better.
}
\begin{tabular}{p{0.15\textwidth} p{0.05\textwidth}<{\centering} p{0.05\textwidth}<{\centering} p{0.05\textwidth}<{\centering} p{0.05\textwidth}<{\centering} p{0.05\textwidth}<{\centering} p{0.05\textwidth}<{\centering} p{0.05\textwidth}<{\centering} p{0.05\textwidth}<{\centering} p{0.05\textwidth}<{\centering} p{0.05\textwidth}<{\centering} p{0.05\textwidth}<{\centering}}
\toprule
& \multicolumn{5}{c}{Mask Segmentation} & \multicolumn{6}{c}{Keypoint Estimation} \\ 
\cmidrule(rl){2-6} \cmidrule(rl){7-12} 
& AP$^\uparrow$ & AP@.50$^\uparrow$ & AP@.60$^\uparrow$ & AP@.70$^\uparrow$ & AP@.80$^\uparrow$ & AP$^\uparrow$ & AP@.50$^\uparrow$ & AP@.60$^\uparrow$ & AP@.70$^\uparrow$ & AP@.80$^\uparrow$ & AR$^\uparrow$ \\ 
\midrule\midrule
Person-in-WiFi \cite{wang2019person} & 0.3800 & 0.9100 & 0.7500 & 0.4000 & 0.0700 & - & - & - & - & - & - \\ 

SiWiS \cite{10.1145/3636534.3690703} & 0.4805 & 0.9452 & 0.8628 & 0.5765 & 0.1055 & 0.3469 & 0.8423 & 0.6595 & 0.3626 & 0.0816 & 0.4559 \\ 

\pname & 0.5097 & 0.9326 & 0.8311 & 0.6614 & 0.2624 & 0.3997 & 0.8995 & 0.7534 & 0.4462 & 0.1140 & 0.5071 \\ 
\bottomrule
\end{tabular}
\label{table:performance}
\end{table*}

% \begin{figure}
%     \centering
%     \includegraphics[width=\linewidth]{picture/performance.pdf}
%     \caption{Qualitative results for keypoint estimation and mask segmentation. \pname demonstrates robust performance compared to vision-based baselines (HRNet and Mask R-CNN) across diverse activities and environments.}
%     \label{fig:performance}
% \end{figure}

\subsection{Evaluation Metrics}

\subsubsection{Mask Segmentation}
We evaluate mask segmentation performance using the standard Intersection over Union (IoU) metric \cite{wang2019person, he2017mask}. IoU quantifies the overlap between the set of predicted pixels \(S_p\) and the set of ground truth pixels \(S_{gt}\). % defined as:
% $
%     \text{IoU} = \frac{\text{area}(S_p \cap S_{gt})}{\text{area}(S_p \cup S_{gt})}
% $,    
% \begin{equation}
%     \text{IoU} = \frac{\text{area}(S_p \cap S_{gt})}{\text{area}(S_p \cup S_{gt})},
% \end{equation}
% where \(S_p \cap S_{gt}\) is the intersection and \(S_p \cup S_{gt}\) is the union of the predicted and ground truth masks. \(\text{area}(\cdot)\) denotes the number of pixels in the area. 
We calculate the average precision (AP) at a given threshold \(\alpha\) as \(\text{AP@}\alpha \!=\! \text{Prob}(\text{IoU} \!\!\ge\!\! \alpha)\)
and 
\(\text{AP} = 0.1 \!\sum_{i=0}^9 \text{AP@}(0.5+0.05i)\).

\subsubsection{Keypoint Estimation}
% We evaluate keypoint estimation performance using two standard metrics: Object Keypoint Similarity (OKS) for overall accuracy and Percentage of Correct Keypoints (PCK) for per-type localization accuracy.

We follow the COCO benchmark protocol  and report Average Precision (AP) and Average Recall (AR) based on Object Keypoint Similarity (OKS) \cite{sun2019deep, xu2022vitpose}:
% The primary evaluation protocal follows the COCO benchmark,
% % \footnote{\url{https://cocodataset.org/\#keypoints-eval}}, 
% which uses Object Keypoint Similarity (OKS) to measure the similarity between predicted poses and ground truth poses:
\begin{equation}
\text{OKS} = \frac{\sum_i \exp(-d_i^2/2s^2k_i^2)\delta(v_i>0)}{\sum_i \delta(v_i > 0)},
\end{equation}
where \(d_i\) is the Euclidean distance between detected keypoints and ground truth, \(v_i\) is the visibility flag, \(s\) is the object scale, and \(k_i\) is a per-keypoint constant that controls falloff. 
%Based on OKS, we report the standard Average Precision (AP) and Average Recall (AR) score \cite{sun2019deep, xu2022vitpose}. 
The main AP metric is computed by averaging AP scores over 10 OKS thresholds: \(\text{AP} = 0.1 \sum_{i=0}^9 \text{AP@}(0.5+0.05i)\), and the main AR metric is calculated based on averaging AR scores at the same OKS thresholds: \(\text{AR} = 0.1 \sum_{i=0}^9 \text{AR@}(0.5+0.05i)\).

To further dissect localization accuracy at the joint level, we employ Percentage of Correct Keypoints (PCK) metric \cite{6380498}:
%To analyze the localization accuracy for each of the \(K\) keypoint types individually, we adopt the PCK metric \cite{6380498}. A prediction is deemed correct if its Euclidean distance to the ground truth is less than a fraction \(\alpha\) of a scale factor \(L_i\). The PCK for keypoint type \(k\) is formulated as:
\begin{equation}
\text{PCK}_k\text{@} \alpha = \frac{\sum_{i=1}^N \mathbb{I}\left( \| \mathbf{p}_{ik} - \mathbf{g}_{ik} \|_2 \leq \alpha \cdot L_i \right) \cdot v_{ik}}{\sum_{i=1}^N v_{ik}} \times 100\%,
\end{equation}
where \(N\) is the total number of human instances, and \(v_{ik}\) is a visibility flag. \(\mathbf{p}_{ik}\) and \(\mathbf{g}_{ik}\) are the predicted and ground truth coordinates of keypoint \(k\) for instance \(i\). The scale factor \(L_i = \sqrt{h(B_{gt_i})^2 + w(B_{gt_i})^2}\) is defined as the diagonal length of the ground truth bounding box \(B_{gt_i}\) of instance \(i\).

\subsection{Evaluation Performance}

We evaluate the performance of \pname by comparing it with two state-of-the-art baselines: Person-in-WiFi \cite{wang2019person} and SiWiS \cite{10.1145/3636534.3690703}. As detailed in Table~\ref{table:performance}, \pname demonstrates superior performance across both mask segmentation and keypoint estimation tasks.

A key observation from our results is that \pname significantly widens its performance gap over the baselines as the evaluation criteria become more stringent. While all methods perform competitively at lower precision thresholds, \pname maintains high fidelity even under demanding requirements. For instance, in mask segmentation, \pname achieves a nearly 150\% improvement in AP@.80 compared to SiWiS. This superior high-precision performance validates the effectiveness of our self-mixing architecture in capturing fine-grained spatial features that are typically lost in traditional CSI-based sensing. For the keypoint estimation task, \pname consistently outperforms the baselines in both Average Precision (AP) and Average Recall (AR). The substantial improvement in these primary metrics indicates that our cross-modal learning framework effectively distills complex body structures from noisy ambient signals, enabling precise localization of individual joints rather than mere presence detection.

% In the mask segmentation task, \pname shows significant advantages, particularly under more stringent evaluation criteria. While achieving competitive results with SiWiS at lower thresholds, \pname noticeably surpasses it as the OKS threshold increases. Specifically, \pname achieves a score of 0.26 for AP@.80, which is nearly 150\% higher than SiWiS. This substantial gain highlights the ability of ARS to achieve a finer-grained spatial resolution, enabling more precise localization of human subjects. This high-precision capability is further corroborated by the results in the keypoint estimation task. \pname consistently outperforms SiWiS across all metrics. For instance, our method improves the primary AP metric from 0.34 to 0.39 and boosts the AR from 0.45 to 0.50. These comprehensive improvements validate the effectiveness of our approach in not only generating accurate body masks but also localizing fine-grained body keypoints. 

To further dissect the system's capabilities, we analyze the Percentage of Correct Keypoints (PCK) across different body parts (Table~\ref{table:keypoint-performance}). Our findings reveal a clear performance hierarchy where torso keypoints, such as shoulders and hips, consistently achieve higher localization accuracy compared to limb extremities like wrists and ankles. This disparity is attributed to two key physical factors: primarily, the larger surface area of the torso provides stronger and more stable RF reflections ; furthermore, core body parts exhibit greater kinematic stability than the rapid and erratic motions typical of hands and feet, which allows our Transformer-based encoder to track them more reliably.

% Next, we conduct a more granular analysis of the localization accuracy for individual keypoints. Table~\ref{table:keypoint-performance} details the Percentage of Correct Keypoints scores for major body parts at various tolerance levels. A key observation is that \pname demonstrates higher accuracy for torso keypoints (shoulders, hips) compared to limb extremities (wrists, ankles). For example, at PCK@.05, \pname achieves scores of 92.09\% for hips and 91.87\% for shoulders, substantially outperforming the 83.67\% for wrists and 79.89\% for ankles. This performance disparity is expected and can be attributed to two physical factors. First, the strength of reflected RF signals is proportional to the surface area of the body part, making the larger torso easier to detect. Second, core body parts exhibit less rapid and erratic motion than the faster-moving extremities, which allows for more stable and accurate localization.

Fig.~\ref{fig:performance} visualizes representative results across various indoor scenes. Even in complex environments with multipath interference, the skeletons and masks generated by \pname remain highly congruent with the vision-based ground truths. These qualitative samples confirm that \pname can serve as a robust, privacy-preserving alternative to camera sensors for fine-grained activity monitoring.

% For a qualitative assessment, we visualize several results from our test set in Fig.~\ref{fig:performance}. These examples compare the outputs of \pname with those from vision-based methods, Mask RCNN \cite{he2017mask} for segmentation and HRNet \cite{sun2019deep} for keypoint estimation. The visualizations demonstrate that \pname delivers robust performance across diverse indoor environments and for various daily activities. 

% A notable advantage of our non-visual approach is its inherent immunity to background clutter that can challenge vision-based systems. For instance, vision-based methods are susceptible to confusing inanimate objects, such as television screens, with people. Our method is completely impervious to such visual-domain errors, ensuring reliable detection regardless of the visual complexity of the scene.

\begin{table}
\scriptsize
\centering
\caption{Percentage of Correct Keypoints (PCK) for different keypoints. The diagonal length of the ground truth bounding box is used as the normalization factor for PCK.}
\begin{tabular}{>{\centering\arraybackslash}p{0.07\textwidth} | >{\centering\arraybackslash}p{0.03\textwidth} >{\centering\arraybackslash}p{0.03\textwidth} >{\centering\arraybackslash}p{0.03\textwidth} >{\centering\arraybackslash}p{0.03\textwidth} >{\centering\arraybackslash}p{0.03\textwidth} >{\centering\arraybackslash}p{0.03\textwidth} >{\centering\arraybackslash}p{0.03\textwidth}}
\toprule
Metric & Nos & Sho & Elb & Wri & Hip & Kne & Ank \\ 
\midrule
\midrule
PCK@.01$^\uparrow$ & 10.46 & 12.41 & 11.67 & 9.75 & 12.30 & 11.13 & 9.04 \\ 
PCK@.02$^\uparrow$ & 34.95 & 40.33 & 39.27 & 32.92 & 40.17 & 37.20 & 31.96 \\ 
PCK@.03$^\uparrow$ & 58.76 & 66.12 & 64.27 & 56.03 & 66.64 & 61.67 & 52.71 \\ 
PCK@.04$^\uparrow$ & 76.23 & 82.90 & 80.68 & 73.09 & 83.50 & 78.94 & 68.71 \\ 
PCK@.05$^\uparrow$ & 86.62 & 91.87 & 89.38 & 83.67 & 92.09 & 88.65 & 79.89 \\ 
\bottomrule
\end{tabular}
\label{table:keypoint-performance}
\end{table}

\subsection{Performance Analysis}

% We now analyze several key factors influencing system performance and provide detailed performance of \pname under various conditions.

We now analyze the key factors influencing the performance of \pname, focusing on how signal duration and detection distance affect sensing accuracy.

\begin{figure}
    \centering
    \includegraphics[width=\linewidth]{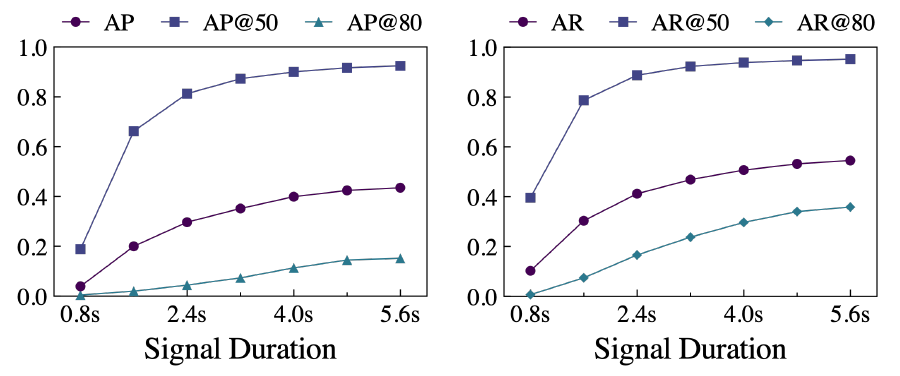}
    \caption{Impact of input signal duration on keypoint estimation performance. System accuracy increases sharply with longer signal durations before plateauing around 4.0 seconds.}
    \label{fig:duration}
\end{figure}

\textbf{Impact of Signal Duration.}
The duration of the input signal sequence is a fundamental factor in capturing the spatio-temporal dependencies of human motion. Our evaluation, conducted across durations ranging from 0.8s to 5.6s, reveals that very brief intervals are insufficient for the model to reconstruct complete poses, primarily because short windows may not capture reflections from all body parts during a kinematic cycle \cite{zhao2018through}. As illustrated in Fig.~\ref{fig:duration}, system accuracy increases sharply as the duration extends toward 4.0s, after which the performance gains begin to saturate. This plateau suggests that a 4.0-second window provides a sufficient temporal receptive field for the Transformer-based encoder to effectively characterize typical human activity patterns. Consequently, we adopt 4.0s as the standard configuration to maintain a balance between sensing fidelity and computational latency.

% The duration of the input signal is a critical factor, as a short time interval may not capture reflections from all body parts, leading to incomplete pose information \cite{zhao2018through}. To quantify this effect, we evaluated the performance of \pname using input signal sequences of varying durations, from 0.8s to 5.6s. 

% The results are illustrated in Fig.~\ref{fig:duration}. Performance is notably poor with a brief 0.8s input but shows a steep improvement as the duration extends to 4.0s. Beyond this point, performance gains begin to saturate, indicating diminishing returns with longer signals. Specifically, increasing the duration from 4.0s to 5.6s yields only a marginal improvement in AP and AR. Therefore, to balance performance with computational efficiency, we adopt a 4.0-second signal duration as the standard configuration for \pname.

\begin{figure}
    \centering
    \includegraphics[width=\linewidth]{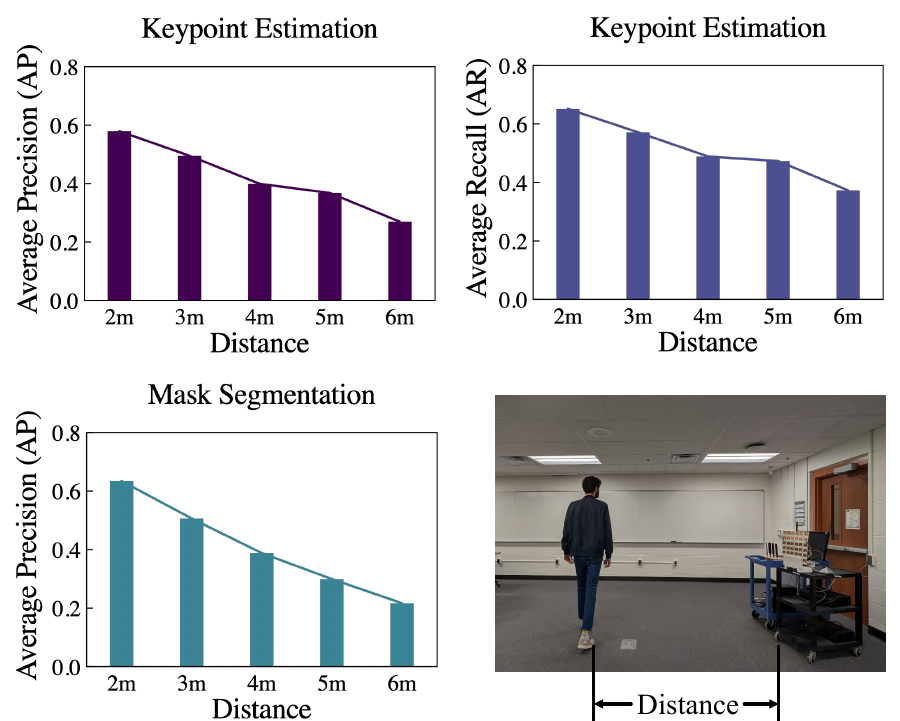}
    \caption{Impact of detection distance on system performance. The bottom-right image depicts the experimental scenario.}
    \label{fig:distance}
\end{figure}

\begin{figure}
    \centering
    \includegraphics[width=\linewidth]{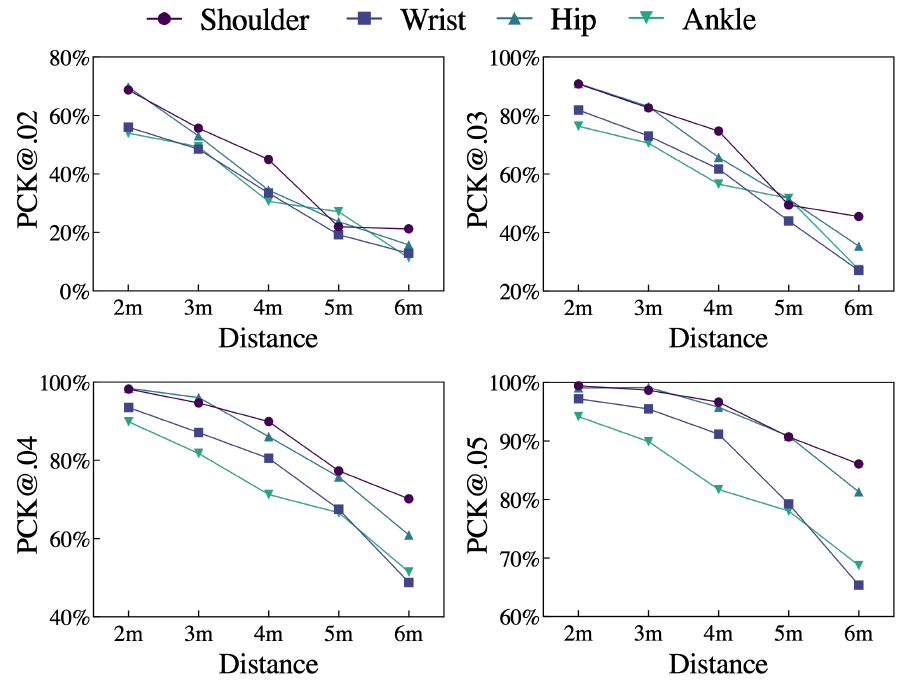}
    \caption{Differential impact of distance on keypoint localization accuracy. Extremities show a more pronounced drop in PCK score compared to the more stable torso keypoints.}
    \label{fig:distance-pck}
\end{figure}

\textbf{Impact of Detection Distance.}
Detection distance presents a significant challenge for RF-based sensing due to the dual physical constraints of signal attenuation and complex multipath propagation. As the subject moves from 2m to 6m away from the \pname device, we observe a downward trend in both keypoint estimation and mask segmentation performance (Fig.~\ref{fig:distance}). This degradation is directly linked to the reduction in effective Signal-to-Noise Ratio (SNR) at longer ranges. A more granular analysis of individual joints further reveals a differential impact of distance: limb extremities exhibit a more pronounced drop in localization accuracy compared to the torso. Specifically, while the larger and more stable shoulder and hip keypoints remain relatively robust, the accuracy for wrists and ankles diminishes sharply as distance increases. This confirms that the weaker reflections and more subtle movements of smaller body parts are the first to be compromised as the signal environment becomes more challenging.

\section{Related Work}

 % RF-based human activity recognition leverages the property that RF signals reflect off the human body to identify various human activities 

\pname is an ISAC technique designed to enable ubiquitous sensing operations. 
It is related to the below areas.

\textbf{ISAC for 5G and Beyond:}
ISAC is a key enabling technology for 5G and beyond \cite{liu2022integrated}.
A variety of techniques have been explored to realize ISAC in future cellular networks, including joint waveform design \cite{zhou2022integrated,xiao2022waveform}, dual-function resource optimization \cite{du2023overview,zhao2022radio,he2023full,dong2022sensing}, localization and tracking applications \cite{zhang2024target}, and efforts toward standardization \cite{luo2025isac}.
\pname is a new approach to ISAC and fundamentally differs from existing efforts.

\textbf{Radar Sensing:}
Radar sensing offers significant advantages in environmental perception and has found widespread applications on mmWave bands \cite{xiao2024survey}.
The rapid advancement of deep learning has enabled a wide range of radar applications, including sleep monitoring \cite{yue2020bodycompass}, gesture recognition \cite{kim2017hand}, radar imaging \cite{adib2015capturing}, and physiological signal monitoring \cite{adib2015smart}.
However, radar sensing typically requires the use of dedicated wideband spectrum. The scarcity of available spectrum below 10 GHz limits the scalability and widespread deployment of radar applications in this frequency range.
\pname addresses the spectrum scarcity issue.

% \textbf{Radar Sensing:}
% Radar sensing offers significant advantages in environmental perception and has found widespread applications on mmWave bands \cite{xiao2024survey}.
% The rapid advancement of deep learning has enabled a wide range of radar applications, including sleep monitoring \cite{yue2020bodycompass, zhao2017learning}, gesture recognition \cite{kim2017hand, lien2016soli, miller2020radsense, skaria2019hand, sun2020real}, radar imaging \cite{adib2015capturing, adib2015multi, bocca2013multiple}, physiological signal monitoring \cite{adib2015smart, chen2020respiration, liu2018non, yue2018extracting}, and object tracking \cite{farella2008interfacing, hsu2017extracting, hsu2019enabling}.
% However, radar sensing typically requires the use of dedicated wideband spectrum. The scarcity of available spectrum below 10 GHz limits the scalability and widespread deployment of radar-based applications in this frequency range.
% \pname addresses the spectrum scarcity issue.

\textbf{Radio Sensing in WiFi:}
Channel State Information (CSI) in WiFi has been extensively studied for human activity detection.
With the rapid advancement of deep learning, a growing number of studies \cite{yang2023sensefi} have developed various DNN-based WiFi CSI sensing applications, such as human activity recognition \cite{yousefi2017survey}, gesture recognition \cite{yang2019learning}, and human pose estimation \cite{zhou2023adapose}.
However, WiFi CSI-based sensing faces fundamental limitations in practical deployments, including the lack of reliable temporal features due to frequency misalignment between transmitting and receiving devices, and the need for multiple coordinated devices.
\pname is designed to complement and overcome these limitations in WiFi CSI-based sensing.

% \textbf{Radio Sensing in WiFi:}
% Channel State Information (CSI) in WiFi has been extensively studied for human activity detection.
% With the rapid advancement of deep learning, a growing number of studies \cite{yang2023sensefi} have developed various DNN-based WiFi CSI sensing applications, such as human activity recognition \cite{yousefi2017survey, zou2018deepsense, li2021two, zou2019wifi, xue2020deepmv, xiao2020deepseg, sheng2020deep, schafer2021human, moshiri2021csi}, human identification \cite{9726794, yang2022efficientfi, yang2022autofi, zhang2018crosssense, gu2021wione, yang2022securesense}, gesture recognition \cite{zhang2018crosssense, zou2018robust, yang2019learning, xiao2019csigan, li2020wihf, wang2022airfi, gu2022wigrunt, zhang2021wifi, zhang2021widar3}, human pose estimation \cite{wang2019can, zhou2023adapose}, and human mask segmentation \cite{wang2019person}.
% However, WiFi CSI-based sensing faces fundamental limitations in practical deployments, including the lack of reliable temporal features due to frequency misalignment between transmitting and receiving devices, and the need for multiple coordinated devices.
% \pname is designed to complement and overcome these limitations in WiFi CSI-based sensing.

\textbf{Human Activity Detection.}
There exists a large body of research on human activity detection, employing various approaches such as radio-based sensing
\cite{wang2019person, zhou2023adapose, zhao2018through, zhao2018rf, zhao2019through}
and camera-based computer vision
\cite{xu2022vitpose, toshev2014deeppose, yang2017learning, sun2019deep, yuan2021hrformer}.
In particular, the computer vision community has made rapid progress in human activity recognition by developing increasingly sophisticated methods, ranging from convolutional neural networks (CNNs) to vision transformers (ViTs), to achieve unprecedented levels of accuracy.
By leveraging these advances in computer vision, \pname can continue to improve its detection performance.

\section{Conclusion}

In this paper, we presented \pname, a novel sensing approach that enables standalone radio devices to leverage over-the-air RF signals from existing communication systems. Through joint hardware–algorithm co-design, \pname uses a self-mixing RF architecture to extract temporal and spatial features of human motion. A cross-modal learning framework further enables vision-supervised training of a radio-based DNN for activity recognition. Prototype experiments demonstrate its effectiveness in human skeleton estimation and body mask segmentation. This work introduces a scalable, spectrum-exempt sensing paradigm for future radio systems.

% In this paper, we presented \pname, a novel approach that enables a standalone radio device to perform sensing by leveraging over-the-air RF signals from an existing communication system.
% \pname is realized through a joint hardware–algorithm co-design. 

% It employs a self-mixing RF architecture to extract both temporal and spatial features of moving objects illuminated by amplified ambient communication RF signals.
% To support real-world applications, we developed a cross-modal learning framework that uses supervision from vision-based models to train a radio-based DNN for human activity recognition. 

% Our prototype implementation and extensive experiments demonstrate the practicality and effectiveness of \pname in two downstream applications: human skeleton estimation and body mask segmentation.
% This work introduces a “spectrum-exempt” sensing paradigm, offering a scalable and occlusion-resilient pathway for next-generation radio sensing systems that can seamlessly coexist with future cellular networks.

\section*{Acknowledgment}
The authors sincerely thank the anonymous reviewers for their constructive comments.
This work was supported in part by NSF Grant ECCS-2434001.

\balance

\clearpage
%% The acknowledgments section 
% \import{section}{acks}

%% the bibliography file.
% \bibliographystyle{ACM-Reference-Format}
\bibliographystyle{ieeetr}
\bibliography{reference}

\end{document}